# Logical inconsistency in combining counterfactual results from non-commutative operations: Deconstructing the GHZ-Bell theorems

Louis Sica
*Chapman University, Orange, CA 92866;
and Inspire Institute Inc., Alexandria, VA 22303, USA*
(May 7, 2012)

E-mail: lousica@jhu.edu

The Greenberger, Horne, Zeilinger (GHZ) theorem is critically important to consideration of the possibility of hidden variables in quantum mechanics. Since it depends on predictions of single sets of measurements on three particles, it eliminates the sampling loophole encountered by the Bell theorem which requires a large number of observations to obtain a small number of useful joint measurements. In evading this problem, the GHZ theorem is believed to have confirmed Bell's historic conclusion that local hidden variables are inconsistent with the results of quantum mechanics. The GHZ theorem depends on predicting the results of sets of measurements of which only one may be performed, i.e., counterfactuals. In the present paper, the non-commutative aspects of these unperformed measurement sequences are critically examined. Three classical examples and two variations on the GHZ construction are analyzed to demonstrate that combined counter factual results of non-commuting operations are in general logically inconsistent with performable measurement sequences that take non-commutation into account. As a consequence, negative conclusions regarding local hidden variables do not follow from the GHZ and Bell theorems as historically reasoned.

## I. INTRODUCTION

The Greenberger, Horn, Zeilinger (GHZ) theorem [1] has achieved a status similar to that of Bell's theorem in its acceptance as an impossibility proof for local hidden variables in quantum mechanics. It has a similarity to Bell's theorem in that it considers a mathematical relation among possible results of alternative measurements that are not all performed. The alternatives consist of exclusive-OR procedures that if performed together are non-commuting, and whose results would be conditional on their order of execution. Thus, if the measurements were all actually performed, their explicit non-commutation would have to be taken into account.

The use of counterfactuals in non-local hidden variables theorems relies on the assumption that counterfactual reasoning is intrinsically sound. But counter examples given in Sec. II B reveal that counterfactual reasoning fails even in the classical domain. It will be argued that such reasoning together with neglect of non-commutation leads to the paradoxical results of the GHZ and Bell theorems. The result is that the discrepancy between quantum mechanical eigenvalues and counterfactual calculations can no longer be taken as proof that local hidden variables are inconsistent with quantum mechanical observations.

A definition of counterfactuals and counter examples showing inconsistencies in their classical use are given in Secs. II A and B. In Sec. III, the accepted interpretation of the GHZ theorem is described following the treatment by Mermin [2], Home [3], Afriat and Selleri [4], and Greenberger [5], but in a manner showing the roll of counter factual reasoning. In sections IV and V, variations on the theorem are considered that also exhibit flaws in counterfactual reasoning. In Sec. VI, analogous inconsistency due to use of counterfactuals in the Bell theorem is outlined. In this case, the logical inconsistency is manifested by violation of the Bell inequality that must be satisfied by cross-correlations of any data sets whatsoever.

## II. COUNTERFACTUAL REASONING

### A. Definition

In the present paper, the term counterfactual will denote a simple one of its various possible usages [6]. If one considers two procedures A and B that do not commute, the result of carrying out a sequence of the two depends on whether A or B is performed first. However, one may consider each procedure in isolation from the other in an exclusive-OR sense. The predicted measurement results of such exclusive-OR procedures that cannot be performed together without taking non-commutation into account, are herein termed counterfactuals. (Since measurement outcomes for commuting procedures have simultaneous existence in quantum mechanics, they are of little concern here.)

### B. Flaws in classical counterfactual reasoning

It has been stated in the context of "no-go" theorems for hidden variables that counterfactual reasoning is used frequently in the classical world without any problem: the principle is logically trustworthy. The author proposes that this belief is in error as will now be shown by classical counter-examples.

We first take note of characteristics of non-commuting operations using a semi-facetious example given several years ago by Leon Cohen in a lecture at the Naval Research Laboratory: putting on shoes and socks. Consider this example from the point of view of counterfactuals. One may put on shoes alone, or socks alone in an exclusive-OR sense, and these acts have perfectly well defined meanings. However, one cannot consider these acts in a logical-AND sense unless non-commutation is taken into account. In that case, putting on socks and then shoes gives a different result from putting on shoes and then socks. Thus, converting the logical-OR case to an AND case in the sense of simultaneous existence, or conversion to commutation without conditionality, makes no physical or logical sense.

A similar situation occurs in the rotation of rigid bodies in three-dimensional space [7]. A rotation of $+90^o$ about the x-axis followed by $+90^o$ about the y-axis produces a different final orientation than if these rotations are carried out in reverse order. Finally, navigation on the surface of the earth is also non-commutative: traveling 100 miles north followed by 100 miles west produces a different final position than the same actions carried out in reverse order due to the definitions of north and west on the spherical surface of the earth. However, in the special case where one begins from a position 50 miles south of the equator the results are the same - but then the operations commute. These examples show that *in general*, one cannot convert counterfactuals of non-commuting operations into commuting ones, or combine them for comparison with the outcomes of the same operations performed non-commutatively, and obtain logically consistency. The situation in quantum mechanics appears to be the same on the basis of analysis to be given in Secs. III - VI.

Interestingly, Griffiths has reached a similar conclusion [8] stating that (counterfactual) results of non-commutative operations "cannot be combined to form a meaningful quantum description" in the consistent histories interpretation of quantum mechanics, and that their joint use is meaningless.

### III. THE GHZ THEOREM

The Pauli spin operators $\sigma_x$, $\sigma_y$, and $\sigma_z$ are used to define three-particle operators $A_1$, $A_2$, and $A_3$:

$$A_1 = \sigma_x^1 \sigma_y^2 \sigma_y^3, \quad A_2 = \sigma_y^1 \sigma_x^2 \sigma_y^3, \quad A_3 = \sigma_y^1 \sigma_y^2 \sigma_x^3, \tag{1}$$

where the superscripts indicate the particle to which the operator is applied. $A_1$, $A_2$, and $A_3$ ultimately act on an entangled state of three spin 1/2 particles. Using the anti-commutation properties of the spin operators,

$$\sigma_i \sigma_j = -\sigma_j \sigma_i, \ i, j = x, y, z, \ i \neq j \tag{2}$$

$$\sigma_i \sigma_j = 1, i = j,$$

and the fact that operators on different particles commute, it is found that $A_1$, $A_2$, and $A_3$ commute. For example, to show that $A_1$ and $A_2$ commute, multiply

$$A_1 A_2 = (\sigma_x^1 \sigma_y^2 \sigma_y^3)(\sigma_y^1 \sigma_x^2 \sigma_y^3) = \sigma_x^1 \sigma_y^1 \sigma_y^2 \sigma_x^2 \sigma_y^3 \sigma_y^3. \tag{3}$$

Then from the anti-commutation property (2):

$$A_1 A_2 = -\sigma_y^1 \sigma_x^1 (-\sigma_x^2 \sigma_y^2) \sigma_y^3 \sigma_y^3 = (\sigma_y^1 \sigma_x^2 \sigma_y^3)(\sigma_x^1 \sigma_y^2 \sigma_y^3) = A_2 A_1. \tag{4}$$

The other commutations may be demonstrated similarly.
One may now consider the product operator $A_1 A_2 A_3$:

$$A_1 A_2 A_3 = (\sigma_x^1 \sigma_y^2 \sigma_y^3)(\sigma_y^1 \sigma_x^2 \sigma_y^3)(\sigma_y^1 \sigma_y^2 \sigma_x^3). \tag{5}$$

Note for later use, that the product (5) may also be written

$$A_1 A_2 A_3 = A^1 A^2 A^3 = (\sigma_x^1 \sigma_y^1 \sigma_y^1)(\sigma_y^2 \sigma_x^2 \sigma_y^2)(\sigma_y^3 \sigma_y^3 \sigma_x^3), \tag{6a}$$

using the definitions

$$A^1 \equiv \sigma_x^1 \sigma_y^1 \sigma_y^1, \quad A^2 \equiv \sigma_y^2 \sigma_x^2 \sigma_y^2, \quad A^3 \equiv \sigma_y^3 \sigma_y^3 \sigma_x^3. \tag{6b}$$

In (5) and (6a), the *sequence* of operations on each individual particle remains the same. Relation (6a) may be simplified by using $\sigma_i^2 = 1$, $i = x, y, z$, to obtain

$$A_1 A_2 A_3 = \sigma_x^1 \sigma_y^2 \sigma_x^2 \sigma_y^2 \sigma_x^3. \tag{7}$$

From (2), this equals

$$A_1 A_2 A_3 = \sigma_x^1 \sigma_y^2 (-\sigma_y^2 \sigma_x^2) \sigma_x^3 = -\sigma_x^1 \sigma_x^2 \sigma_x^3 = -A_4, \tag{8a}$$

where $A_4 = \sigma_x^1 \sigma_x^2 \sigma_x^3$. $A_4$ commutes with $A_1$, $A_2$, and $A_3$, and

$$A_1 A_2 A_3 A_4 = -\sigma_x^1 \sigma_x^2 \sigma_x^3 \sigma_x^1 \sigma_x^2 \sigma_x^3 = -1, \tag{8b}$$

for any state of the three particles.

The GHZ theorem depends on the above state-independent properties of the $A_i$, and further consequences that follow from the assumption that the three particles are described by the entangled state

$$|\psi\rangle = \frac{1}{\sqrt{2}} \left( |\alpha_1\rangle|\alpha_2\rangle|\alpha_3\rangle - |\beta_1\rangle|\beta_2\rangle|\beta_3\rangle \right). \tag{9}$$

In (9), $|\alpha_i\rangle$ and $|\beta_i\rangle$, for particles $i = 1, 2, 3$ designate the eigenkets of $\sigma_z$ with eigenvalues +1 and -1, respectively. By using the well known relations [9]

$$\sigma_x|\alpha\rangle = |\beta\rangle, \sigma_x|\beta\rangle = |\alpha\rangle, \sigma_y|\alpha\rangle = i|\beta\rangle, \sigma_y|\beta\rangle = -i|\alpha\rangle, \tag{10}$$

the operation of $A_1$, $A_2$, and $A_3$ on $|\psi\rangle$ yields

$$A_i|\psi\rangle = +1|\psi\rangle, \ i = 1, 2, 3. \tag{11a}$$

Then from (8a), which follows from the spin anti-commutation relations,

$$A_4|\psi\rangle = -1|\psi\rangle. \tag{11b}$$

Thus, $|\psi\rangle$ is a common eigenstate of all the $A_i$. This implies that the $A_i$'s $i = 1, \ldots, 4$ are simultaneously measurable. Thus, for example, the same value of $A_1$ occurs at each occurrence in the measurement sequence $A_1 A_2 A_1$, since the operators commute. However, a measurement of an $A_i$ must be made in such a way that only the product of the spin values and not their individual values are determined [5]. Otherwise, the states produced by the $A_i$'s after measurement operations would not equal the state $|\psi\rangle$, an entangled state in which no specific values of spin components are specified. Consistent with this, is the fact that $|\psi\rangle$ is not an eigenstate of $A^1$, $A^2$ or $A^3$ defined in (6b). Thus, one cannot choose to perform the sequence of measurements prescribed by $A^i$ on particle $i$, to obtain an eigenvalue of $|\psi\rangle$, though $A^1$, $A^2$, $A^3$ commute with each other and $A_4$.

The argument that (11a,b) is not consistent with local realism, i.e., the existence of hidden variables or pre-existing values for measurement outcomes, is as follows. If local hidden variables supplementing the information in $|\psi\rangle$ are assumed to exist, their values would determine the components of spin found in measurements performed on the three particles. If the particles were separated after the formation of $|\psi\rangle$, under the assumption of locality the measured value obtained for any selected particle spin component would be independent of the *choice* of component measured on any other distant particle. Thus, the value of $\sigma_y$ obtained for particle 3 would be independent of whether one chose to measure $\sigma_y$ or $\sigma_x$ on particle 2. In view of (11a), one could measure either $A_1$, $A_2$ or $A_3$ on $|\psi\rangle$ to obtain

$$m_x^1 m_y^2 m_y^3 = 1, \ m_y^1 m_x^2 m_y^3 = 1, \ \text{or} \ m_y^1 m_y^2 m_x^3 = 1, \tag{12a, b, c}$$

respectively, where the m's, each equal to $\pm 1$, denote numerical values of the (counterfactual) measurements. But values of the same symbol occurring in different relations (12a-c) must be the same based on the assumption that the particles do not interact after separation, and on the assumption that the values result from initial conditions determining the exclusive-OR measurement values of $A_1$, $A_2$ and $A_3$. Thus, if (12a) is measured, the value of $m_y^3$ must be the same as if one had chosen to measure (12b) instead, and the value of $m_y^1$ occurring in (12b) must be the same as that occurring in (12c), since changing the measurement on the other two particles could have no effect on it after the particles had separated. (Note, as mentioned above, that only one of the $A_i$'s can be measured so that the measurement choice is an exclusive-OR choice.) Based on this counter factual reasoning, one can combine the exclusive-OR results to obtain the product of (12a), (12b), and (12c):

$$(m_x^1 m_y^2 m_y^3)(m_y^1 m_x^2 m_y^3)(m_y^1 m_y^2 m_x^3) = 1. \tag{13}$$

Since the values for each of the two recurrences of $m_y^i$, $i = 1, 2, 3$ are equal in (12a, b, c),

$$m_x^1 m_x^2 m_x^3 = 1. \tag{14a}$$

But from (11b) which takes non-commutation into account to predict the results of real measurements,

$$m_x^1 m_x^2 m_x^3 = -1. \tag{14b}$$

Hence, based on the use of counter factuals, the set of predictions of quantum mechanics appears to be inconsistent with the existence of preexisting or predetermined values for local variables.

Similar inconsistency may be demonstrated in a more succinct and state independent manner by applying counterfactuals to (8b) to obtain +1 rather than the quantum result -1:

$$(m_x^1 m_y^2 m_y^3)(m_y^1 m_x^2 m_y^3)(m_y^1 m_y^2 m_x^3)(m_x^1 m_x^2 m_x^3) = (m_x^1 m_y^1 m_y^1 m_x^1)(m_y^2 m_x^2 m_y^2 m_x^2)(m_y^3 m_y^3 m_x^3 m_x^3) = 1, \tag{15}$$

since each spin operator on the left side of (8b), and corresponding measurement counterfactual in (15) occurs twice, with counterfactuals equal to $\pm 1$. This state-independent disagreement between the counterfactual and quantum results is a version of the Kochen-Specker theorem [2]. The reasoning is again based on the use of values resulting from measurement operations of which only one may be performed in an exclusive-OR sense, but which are then combined in an AND sense. If this reasoning were sound and the effects of non-commutation could be neglected in quantum mechanics, it could be argued that the inconsistent results of state independent (8b) and counterfactual counterpart (15), are evidence of non-locality or the non-existence of pre-existing measurement values or causal processes. However as seen in Sec. II B, such reasoning does not hold classically, and from the present analysis does not appear to hold in quantum mechanics either.

## IV. GHZ-LIKE RESULT FOR NON-ENTANGLED PARTICLES

While the GHZ results (11) and (14) appear to depend on entangled state (9), since each of $A_1$, $A_2$, $A_3$, has eigenvalue 1 when operating on (9), important aspects of the algebra of the situation are the same when (9) is replaced by the product state

$$|\psi_2\rangle = \frac{1}{2^{3/2}}(|\alpha_1\rangle - |\beta_1\rangle)(|\alpha_2\rangle - |\beta_2\rangle)(|\alpha_3\rangle - |\beta_3\rangle). \tag{16}$$

Using (10) for example, one finds that

$$A_1|\psi_2\rangle = \sigma_x^1 \sigma_y^2 \sigma_y^3 |\psi_2\rangle = \frac{1}{2^{3/2}} \sigma_x^1(|\alpha_1\rangle - |\beta_1\rangle) \sigma_y^2(|\alpha_2\rangle - |\beta_2\rangle) \sigma_y^3(|\alpha_3\rangle - |\beta_3\rangle)$$
$$= \frac{1}{2^{3/2}}(|\beta_1\rangle - |\alpha_1\rangle) i(|\beta_1\rangle + |\alpha_1\rangle) i(|\beta_3\rangle + |\alpha_3\rangle) \tag{17}$$

so that $|\psi_2\rangle$ is not an eigenstate of $A_1$, $A_2$, or $A_3$ acting alone. However, it is an eigenstate of the product (7) of the three operators $A_1 A_2 A_3 = \sigma_x^1 \sigma_y^2 \sigma_x^2 \sigma_y^2 \sigma_x^3$, and of $A^1$, $A^2$, $A^3$, and $A_4$ acting alone. This is due to the fact that the operation of $\sigma_x^1$ and $\sigma_x^3$ reverse the sign of the first and third factors of (16) respectively, while $\sigma_y^2 \sigma_x^2 \sigma_y^2$ leaves the sign of the middle factor unchanged. Groupings of individual particle spin operators to form $A^1$, $A^2$, $A^3$, and $A_4$, all of which commute, now have a common eigenstate in $|\psi_2\rangle$. Since, each of the factors of (16) corresponds to an eigenvalue -1 for $\sigma_x$, pre-existing spin-values exist for each of the three particles. As there is no quantum mechanical entanglement, one can choose to carry out measurements on any particle without affecting the others. However, one still finds that

$$A_1 A_2 A_3 |\psi_2\rangle = A^1 A^2 A^3 |\psi_2\rangle = \sigma_x^1 (\sigma_y^2 \sigma_x^2 \sigma_y^2) \sigma_x^3 |\psi_2\rangle = 1|\psi_2\rangle, \tag{18}$$

and

$$A_4 |\psi_2\rangle = \sigma_x^1 \sigma_x^2 \sigma_x^3 |\psi_2\rangle = -1|\psi_2\rangle. \tag{19}$$

One now applies counterfactual reasoning to (18) and (19) analogously to Sec. III. Measurements of $A^1 = \sigma_x^1$ and $A^3 = \sigma_x^3$ to obtain $m_x^1$ and $m_x^3$ are multiplied by counterfactual values for $\sigma_y^2 \sigma_x^2 \sigma_y^2$, i.e., $m_y^2 m_x^2 m_y^2$, similarly to the use of simultaneously un-measureable values for $A_1$, $A_2$, $A_3$ as in Sec. III. The counterfactual product corresponding to (18) is thus

$$m_x^1(m_y^2 m_x^2 m_y^2)m_x^3 = m_x^1 m_x^2 m_x^3 = 1, \tag{20}$$

while by taking non-commutation into account, the quantum mechanical result from (19) is

$$m_x^1 m_x^2 m_x^3 = -1. \tag{21}$$

Thus, in spite of the fact that (16) is a product state specifying pre-existing spin measurement values, the same kind of result follows as in Sec. III when non-commutation is neglected in counter factuals multiplied together.

## V. GHZ-LIKE RESULT FOR MEASUREMENT ON A SINGLE PARTICLE

From the pattern of non-commutative operator actions on particle 2, it is evident that a similar result can be obtained by using a single particle alone, along with the operators previously applied to particle 2. Define a state $|\psi_3\rangle$ equal to

$$|\psi_3\rangle = \frac{1}{\sqrt{2}}(|\alpha\rangle - \beta). \tag{22}$$

Then consider the effect of commuting operators $\sigma_x$ and $\sigma_y \sigma_x \sigma_y$ on (22):

$$\sigma_y \sigma_x \sigma_y |\psi_3\rangle = \sigma_y \sigma_x \sigma_y \frac{1}{\sqrt{2}}(|\alpha\rangle - |\beta\rangle) = +1|\psi_3\rangle, \tag{23a}$$

$$\sigma_x |\psi_3\rangle = \sigma_x \frac{1}{\sqrt{2}}(|\alpha\rangle - |\beta\rangle) = \frac{1}{\sqrt{2}}(|\beta\rangle - |\alpha\rangle) = -|\psi_3\rangle, \tag{23b}$$

with

$$(\sigma_y \sigma_x \sigma_y)\sigma_x = \sigma_x(\sigma_y \sigma_x \sigma_y) = -1. \tag{23c}$$

The counterfactual version of (23a), with $\sigma_y$ and $\sigma_x$ replaced by inferred counter factual values determined from initial conditions, yields

$$m_y m_x m_y = m_x (m_y)^2 = m_x = 1, \tag{23d}$$

whereas quantum mechanics (23b) predicts $m_x = -1$. From this, as well as the previous examples, a logically sufficient explanation of the arithmetic paradox of the GHZ theorem is the fact that one cannot replace non-commuting operator products with products of exclusive-OR counterfactual initial values.

The examples in Secs. III, IV and V, with and without entanglement, show that one cannot combine exclusive-OR results of non-commutative operations as if they were results of commutative AND operations, and expect that the numerical values will be consistent with those obtained from actual non-commutative measurement sequences. The output of each non-commutative operation depends on the previous outcome, so that *even if a counterfactual construction makes logical sense in and of itself, a different numerical value will be produced than when the real experimental operations are performed non-commutatively.*

# VI. BELL'S THEOREM

The present analysis would not be complete without some discussion of Bell's theorem, which also depends on counterfactual reasoning applied to non-commuting operations. The author has pointed out the impact of non-commutation on the logical ingredients of the theorem in previous publications [10], but a brief recapitulation in the specific context of flawed counterfactual reasoning is appropriate here. It is easy to show that the inequality that Bell derived is a mathematical identity: it is identically satisfied by cross-correlations of any three or four (as appropriate to the number of variables treated) data sets consisting of $\pm 1's$. This fact depends *only* on the assumed existence of the data sets, and is independent of any other property such as their origin in random or deterministic processes, locality, or commutation or non-commutation of measurement processes. However, Bell derived the inequality under less general assumptions: that the correlations result from a process that is stationary in second order [11] (all correlations are given by the same function of the difference of instrument settings), all measurements are commutative or at least counterfactually defined, and locality. Of course, the inequality must be satisfied by the cross-correlations of the data sets of such special processes also, and by their resulting single correlation function evaluated at differences of pairs of instrument settings.

However, in the quantum mechanical two particle experiments to which this inequality has been applied, consideration of more than one measurement per particle implies that non-commutation must be taken into account. While the first two measurements, one per particle, are commutative, any additional actual measurements are non-commutative and conditional on the first. Ideally, they would come from apparatuses in tandem with those carrying out initial measurements, with the whole assembly operating in a retrodictive mode. However, Bell did not consider real measurements [12], but a counterfactual alternative for values beyond the first two, with correlations between real and counterfactual outcomes given by the single correlation function that characterizes his assumption of stationary stochastic processes. Thus, counterfactual outcomes were used in place of the results of sequential non-commuting measurement operations. The resulting mathematical inconsistency is registered as violation of the Bell inequality, which must be satisfied by cross-correlations of any data sets whatsoever.

In practice, the data from experiments have not been cross-correlated, and each pair of correlations is derived from an independent experimental run, as is consistent with Bell's stationary stochastic model. If the underlying process were second-order stationary as assumed by Bell, there would only be one correlation functional form to determine, and ensemble averaged cross-correlations would yield the same function as that measured in independent runs. The Bell inequality would be satisfied by correlation functions measured in independent runs up to small statistical fluctuations. The violation of the inequality by such experimentally confirmed Bell cosine correlation functions proves that the underlying process is not statistically stationary. This is consistent with the non-commutative process described by quantum mechanics, and implies different correlation functional forms between the different variables in a non-commutative sequence. It is again consistent with the logical proposition that counterfactuals of non-commutative processes are inconsistent with measurement predictions that properly take non-commutation into account.

# VII. CONCLUSION

Counterfactual reasoning, both classically and quantum mechanically, cannot be expected to produce results that are logically consistent with those based on taking non-commuting operations properly into account. The no-hidden-variables theorems of quantum mechanics described above depend on such counterfactual reasoning. The narrative accompanying these theorems is that counterfactual reasoning involving non-commuting variables is logically sound, so it is appropriate to attribute the peculiarly inconsistent results that follow to non-locality, or the non-existence of hidden variables or pre-existing values. But if the use of counterfactual reasoning in no-hidden-variables theorems is flawed both classically and quantum mechanically as shown above by counter examples, the universally accepted conclusions that hidden variables are inconsistent with quantum mechanics does not follow. That said, the existence of defects in no-hidden-variables theorems does not, in and of itself, imply that local hidden variables exist.

# ACKNOWLEDGEMENTS

I would like to thank Mike Steiner of Inspire Institute for useful critical comments on the manuscript, and Armen Gulian and Joe Foreman of the quantum group at Chapman University East for useful discussions relating to the presentation of the material.

______________________________________________________________________

[1] D. M. Greenberger, M. Horne, and Z. Zeilinger, in *Bell's Theorem Quantum Theory and Conceptions of the Universe*, edited by M. Kafatos (Kluwer, Dordrecht, 1989), p. 73.
[2] N. David Mermin, Phys. Rev. Lett., 65, 3373 (1990).
[3] D. Home, *Conceptual Foundations of Quantum Physics* (Plenum Press, New York, 1997), p. 234.
[4] A. Afriat and F. Selleri, *The Einstein, Podolsky, and Rosen Paradox* (Plenum Press, New York, 1999), p.121.
[5] D. M. Greenberger, *Compendium of Quantum Physics*, edited by D. Greenberger, K. Hentschel, F. Weinert (Springer, Dordrecht, Heidelberg, London, New York, 2009), p. 258.
[6] L. Vaidman, *Compendium of Quantum Physics*, edited by D. Greenberger, K. Hentschel, F. Weinert, (Springer, Dordrecht, Heidelberg, London, New York, 2009), p. 132.
[7] H. Goldstein, *Classical Mechanics* (Addison-Wesley, Reading, Mass., 1980), p. 148.
[8] R. B. Griffiths, *Compendium of Quantum Physics*, edited by D. Greenberger, K. Hentschel, F. Weinert (Springer, Dordrecht, Heidelberg, London, New York, 2009), p. 117.
[9] F. Mandl, *Quantum Mechanics* (John Wiley and Sons, Chichester, England, 1992), Chap. 5.
[10] L. Sica, Opt. Commun. 170, 55 (1999) ; 170, 61 (1999); J. Mod. Opt. 59, No.15-17, 2465 (2003).
[11] A. Papoulis and S. U. Pillai, *Probability, Random Variables, and Stochastic Processes* (McGraw-Hill, New York, N. Y., 2002), p. 387.
[12] J. S. Bell, *Speakable and unspeakable in quantum mechanics* (Cambridge University Press, Cambridge, 1987), p. 65.